\def   \ni {\noindent}
\def   \bsk {\vskip 15truept}

\documentstyle[epsfig]{article}
\begin{document}

\font\affiliation=cmssi10
\font\author=cmss10
\font\caption=cmr8
\font\references=cmr8
\font\title=cmssbx10 scaled\magstep2
\def\ref{\par\noindent\hangindent 15pt}
\null

\title{\ni Thermal and Viscous Instability
             of Accretion Disc in AGN 
 }
                                
\bsk \bsk
\author{\ni A.~Janiuk $^{1}$, B.~Czerny $^{1}$ }

\bsk
\affiliation{1) N. Copernicus Astronomical Centre,  Bartycka 18, 00-716, Warsaw, Poland
}       
\bsk
\baselineskip = 12pt

\abstract{\ni

The observed optical/UV spectra of most Seyfert galaxies are much redder than
expected from a stationary accretion disk. Two explanations of this efect are
probable: (i) the observed spectrum is strongly contaminated by starlight (ii) 
the accretion disk is not stationary in these objects.

 The standard accretion discs are
 known to be thermally and viscously unstable over a certain range of 
temperatures. In the inner disc regions there may develop radiation pressure
 driven instability, which is possibly related to the rapid variability 
detected in AGNs in the UV range. In the outer disc develops the ionization 
instability, similar to that in the cataclysmic variables, but
operating on much longer timescales. Due to this process the spectrum of the
 accretion disc differs from that of a stationary one.

We examine the accretion disc vertical structure model in order to determine 
the range of the ionization instability. We derive the radial dependence
 of the accretion rate in a non-stationary disc between outbursts, 
and we calculate the effective
 temperature profile and the corresponding spectral slope.  

The predicted shape of the spectrum is different from a stationary case but
it does not seem to reproduce the observed spectra of Seyfert galaxies.
We therefore suggest that the starlight contribution is most probably
responsible for the spectral shape in red Seyfert galaxies and its contribution
extends to shorter wavelengths than usually adopted.

}                                                    
\bsk
              
\section{Introduction}

Accretion disc in Active Galactic Nuclei are subject to thermal - viscous 
instabilities which operate in the inner disc regions. The radiation pressure 
driven instability (Lightman \& Eardley 1974) is developed in the models, 
which assume the viscous torque proportional to the total pressure.
It operates in the innermost disc part, while further away, where the disc is 
gas pressure dominated, there exists a partial ionization zone. Therefore these
 regions are unstable due to the same mechanism as in cataclysmic variables
(e.g. Smak 1984, Lasota et al. 1995), but in much longer time scales.
Finally, the outer parts of the disc may be gravitationally unstable 
(Hure 1998).

Here we consider only the ionization instability. Its radial extension depends
on several parameters, like the mass of a central black hole, viscosity 
parameter $\alpha$ and external acceretion rate $\dot M_{ext}$. Using the 
parametrization given in Siemiginowska et al. (1996) we estimate for the inner
 and outer radii of the unstable zone that 
$R_{in} \sim 250 R_{schw}$ and $R_{out} \sim 1500 R_{schw}$ for 
$M = 10^{8} M_{\odot}$, $\alpha = 0.03$ and $\dot M_{ext} = 0.1\dot M_{Edd}$.

At any given radius the local thermal equilibrium may be represented by the 
chracteristic shape of the {\bf S} curve on the $T_{eff} - \Sigma$ 
(or $\dot M - \Sigma$) plane. The upper and lower branches of the positive 
slope describe the  stable configurations
and the middle part of the  {\bf S} curve 
describes the  unstable solution. 
When the external accretion rate corresponds to the unstable branch the disk
undergoes oscillations between the upper hot and the lower cold branch. 
Faint phase corresponds to the slow accumulation of the material and increase
of the disk surface density; the accretion rate and surface density 
corresponding to the upper end of the low branch are characteristic for this 
phase. 

\bsk
              
\section{Vertical structure model}

The basic equations that describe the disc vertical structure are the 
equation of 
viscous energy dissipation, the hydrostatic equilibrium, 
and equation of energy transfer. The last equation takes into account the 
presence of convection which carries non-negligible fraction of energy.
The frequency-averaged opacity $\kappa$ (Rosseland mean) 
includes the electron scattering as
well as the required free-free and bound-free transitions. The tables are
from Alexander, Johnson \& Rypma (1983) 
for log $T <3.8$, from Seaton et al. (1994) for  log $T >4.0$ and the value
of opacity is 
interpolated between these two tables for intermediate values of the
temperature, as in R\' o\. za\' nska et al. (1999).  

The temperature dependence of the Rosseland mean opacity changes for different 
temperature ranges: for $T < 10^{3}$ K the power law function is: 
$\kappa \sim \rho T^{-5}$, for $T > 10^{4}$ K the power law index is equal to 
-3.5, while at the intermediate temperature range, corresponding to partial 
hydrogen ionization, the trend is opposite and  power law index is about 
5 - 10 (see Kato et al. 1999).
The opacity changes affect the cooling rate in the disc, which is important 
in determining the equilibrium solutions. Moreover, the presence of $H^{-}$
ions causes a large reduction of radiative transport efficiency and 
therefore excited convection results in higher opacity.

\section{Results}

In Figure 1. we plot local stability curves for several different radii. The 
critical turning points on each curve, indicated as $a_{1}$, $a_{2}$ and 
$a_{3}$, represent the maximum surface densities achieved during the faint
phase of the ionization
 instability cycle. The critical mass transfer rate is 
$\dot M_{crit}(r)\equiv \dot M (\Sigma_{max}(r))$.    

\begin{figure}
\centerline{\psfig{file=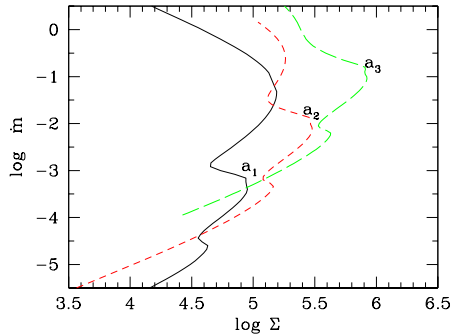, width=7cm}}
\caption{FIGURE 1. The theoretical stability curves for $M=10^{8} M_{\odot}$,
 $\alpha = 0.03$ and three values of
radius $r$: $r = 100 R_{Schw}$ (solid line), $r = 350 R_{Schw}$ (dashed
line),  and $r = 1000 R_{Schw}$ (long 
dashed line). The accretion rate is given in Eddington units.
}
\end{figure}


In Figure 2. we show the radial dependence of accretion rate in the faint
phase, derived from 
our vetical structure model. The accretion rate is in Eddington units and the 
radius is in Schwarzschild radii. This dependence is well fitted with the 
power law function: $\dot m(r) \sim r^{2.3}$.

\begin{figure}
\centerline{\psfig{file=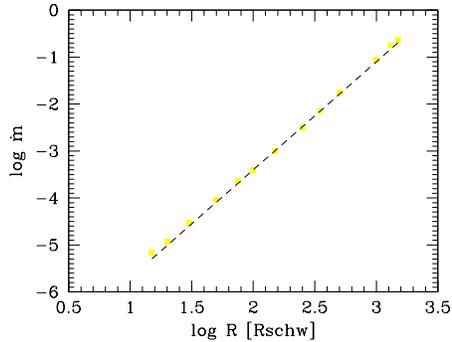, width=7cm}}
\caption{FIGURE 2. The radial dependence of accretion rate  
(vertical structure model parameters: $M=10^{8} M_{\odot}$,
 $\alpha = 0.03$) . The accretion rate is given in Eddington units.
The dashed line represents the power law function $\dot m(r) \sim r^{2.3}$.
}
\end{figure}


The effective temperature of the accretion disc scales with radius as 
$T(r) \sim (\dot M(r)/r^{3})$ and for the stationary accretion disc this 
relation gives $T(r) \sim r^{-0.75}$, while in our model we obtain
$T(r) \sim r^{-0.175}$.
The approximate spectral slope is given by the formula 
$F_{\nu} \sim \nu^{3-2/p}$, where $p$ is the index in the
radial dependence of the temperature (Kato et al. 1999).
In the first case it results in $F_{\nu} \sim \nu^{1/3}$.
In the latter case
it becomes  $F_{\nu} \sim \nu^{-8.4}$, however as the outer disc temperature
is not very much smaller than the inner disc temperature, 
the resulting spectrum looks practically like a single black body.
 None of these, however, is observed in 
the Seyfert galaxies, for which the optical/UV spectral indices are in the 
range from -1.9 to -1.2 (e.g. Edelson \& Malkan 1986).
For NGC 7469 the  spectral slope in the optical band is $\sim -1.69$ 
(see also Nandra et al. 1998),
 while for the faint state of  NGC 4151 it is $\sim -1.7$ (Lyutyi \& 
Doroshkievich, 2000).   

Therefore our conclusion is, that even if the instability processes may
 affect the emitted optical spectral slope, nevertheless the non-negligible 
effect of starlight component should be taken into account. 

In Figure 3. we show the example of the starlight spectrum, taken to be the 
spectrum of the nuclear bulge of M31 galaxy, as in Kuraszkiewicz et al. (1997).
It is  compared with the 
power law spectrum of the index -1.9 ( e.g. NGC 4051, NGC 3516, Mkn 231).
We conclude, that the starlight contribution may be responsible for this spectral slope,
however it must extend to shorter wavelengths. Clearly
more young, blue stars are needed in this case to account for this 
shape of the spectrum.

\begin{figure}
\centerline{\psfig{file=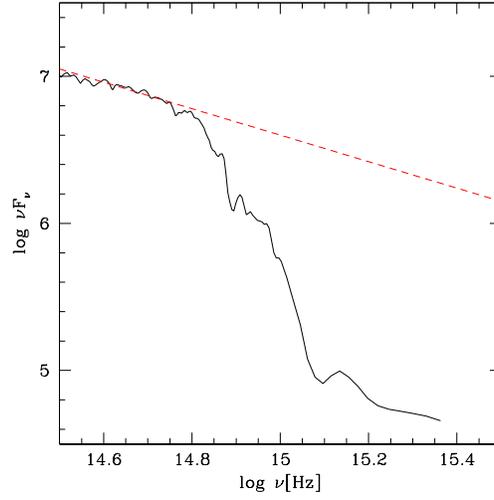, width=7cm}}
\caption{FIGURE 3. The starlight spectrum (solid line) compared with the
power law $\nu F_{\nu} \sim \nu^{-0.9}$ (dashed line). 
}
\end{figure}


\bsk
\baselineskip = 12pt

{\references \ni REFERENCES
\bsk

\ref Alexander D.R., Johnson H.R., Rypma R.L., 1983, ApJ, 272, 773

\ref Done C., et al., 1990, MNRAS 243, 713

\ref Edelson R.A., Malkan M.A., 1986, ApJ, 308, 59

\ref Hure J.M., 1998, A\&A, 337, 625

\ref Kato S., Fukue J., Mineshige S., ``Black hole accretion discs'', 1999, 
Kyoto Univ. Press 

\ref Kuraszkiewicz J., Loska Z., Czerny B., 1997, Acta Astron., 47, 263

\ref Lasota J.P., Hameury J.M., Hure J.M., A\&A, 1995, 302, L29

\ref Lightman A.P., Eardley D.M., 1974, ApJ, 187, L1 

\ref Lyutyi \& Doroshkievich, 2000, JENAM Conf., Moscow

\ref Nandra K., et al., 1998, ApJ, 505, 594

\ref R\' o\. za\' nska A., Czerny B., \. Zycki P.T., Pojma\' nski G., 1999, MNRAS, 305, 481

\ref Seaton M.J., Yan Y., Mihalas D., Pradhan A.K., 1994, MNRAS, 266, 805

\ref Siemiginowska A., Czerny B., Kostyunin V., 1996, ApJ, 458, 491

\ref Smak J.I., 1984, Acta Astron., 34, 161
}

\end{document}